\documentclass{JHEP3}
\usepackage[dvips]{graphicx}
\usepackage{amsmath,amssymb,bm}

\def\imo{i}

\def\a{\widetilde{\alpha}}
\def\imo{i}

\def\Re#1{\mathrm{Re}(#1)}
\def\Im#1{\mathrm{Im}(#1)}

\title{\textbf{Quasinormal modes of Gauss-Bonnet-AdS black holes: towards holographic description of finite coupling}}
\author{\textbf{R. A. Konoplya}\\
Theoretical Astrophysics, Eberhard-Karls University of T\"ubingen, T\"ubingen 72076, Germany\\and\\Institute  of  Physics  and  Research  Centre  of  Theoretical  Physics  and  Astrophysics, Faculty  of  Philosophy  and  Science,  Silesian  University  in  Opava,  Opava,  Czech  Republic}
\author{\textbf{A. Zhidenko}\\
Centro de Matem\'atica, Computa\c{c}\~ao e Cogni\c{c}\~ao, Universidade Federal do ABC (UFABC), Rua Aboli\c{c}\~ao, CEP: 09210-180, Santo Andr\'e, SP, Brazil}
\date{}

\preprint{}

\abstract{
Here we shall show that there is no other instability for the Einstein-Gauss-Bonnet-anti-de Sitter (AdS) black holes, than the eikonal one and consider the features of the quasinormal spectrum in the stability sector in detail. The obtained quasinormal spectrum consists from the two essentially different types of modes: perturbative and non-perturbative in the Gauss-Bonnet coupling $\alpha$. The sound and hydrodynamic modes of the perturbative branch can be expressed through their Schwazrschild-AdS limits by adding a linear in $\alpha$ correction to the damping rates: $\omega \approx \Re{\omega_{SAdS}} - \Im{\omega_{SAdS}} (1 - \alpha \cdot ((D+1) (D-4) /2 R^2)) \imo$, where $R$ is the AdS radius. The non-perturbative branch of modes consists of purely imaginary modes, whose damping rates unboundedly increase when $\alpha$ goes to zero. When the black hole radius is much larger than the anti-de Sitter radius $R$, the regime of the black hole with planar horizon (black brane) is reproduced. If the Gauss-Bonnet coupling $\alpha$ (or used in holography $\lambda_{GB}$) is not small enough, then the black holes and branes suffer from the instability, so that the holographic interpretation of perturbation of such black holes becomes questionable, as, for example, the claimed viscosity bound violation in the higher derivative gravity. For example, $D=5$ black brane is unstable at $|\lambda_{GB}|>1/8$ and has anomalously large relaxation time when approaching the threshold of instability.}

\begin{document}

\section{Introduction}

Quasinormal modes of asymptotically anti-de Sitter black holes play an essential role in the description of strongly coupled processes in the dual conformal field theory via the AdS/CFT correspondence \cite{Maldacena:1997re}. The poles of the retarded Green functions in the strongly coupled Conformal Field Theory in $D$-dimensions coincide with the proper oscillation frequencies (called \emph{quasinormal modes}) of large (in comparison with the AdS radius) $(D+1)$-dimensional asymptotically anti-de Sitter black holes \cite{Birmingham:2001pj}. The damping rate of the dominant quasinormal mode determines the timescale at which a black hole or quantum system in the dual field theory  approaches equilibrium \cite{Son:2007vk}.

In 2005 Kovtun, Son, and Starinets showed \cite{Kovtun:2004de} that the AdS/CFT correspondence predicts the following universal behavior of quark-gluon plasmas when modeled in the Conformal Field Theory through various gravitational backgrounds:
\begin{equation}\label{qq}
\frac{\eta}{s} \approx \frac{\hbar}{4 \pi k},
\end{equation}
where $\eta$ is the shear viscosity, $s$ is volume density of entropy. They also suggested that  (\ref{qq}) determines the lower bound
on the shear viscocity/entropy density ratio. Soon the theoretical prediction (\ref{qq}) was confirmed through comparisons with the data obtained when observing quark-gluon plasmas at RHIC \cite{Luzum:2008cw}.

However, the above results were obtained at the assumption that the dual field theory has large `t Hooft coupling $\lambda$ \cite{Son:2007vk}.
The regime of weak coupling can be described by the kinetic theory, while the transition from weak to strong coupling, i.~e., the regime of intermediate coupling is not yet understood. Recently, an approach allowing one to obtain corrections to the regime of infinite coupling has been  suggested  in \cite{Waeber:2015oka,Grozdanov:2016vgg}. There it is shown that higher curvature corrections to the gravitational action, such as the Gauss-Bonnet, Lovelock, $R^4$ or other terms, may give us a hint on what happens in the regime of intermediate coupling.

In \cite{Grozdanov:2016vgg,Grozdanov:2016fkt,Grozdanov:2015asa} quasinormal modes of black branes in various theories with higher curvature corrections were investigated and several new interesting findings were reported. The main striking difference of the black-hole spectrum in presence of higher curvature corrections is  appearance of a new, purely imaginary (i.~e., non-oscillating), non-perturbative (in terms of the Gauss-Bonnet coupling constant $\alpha$) modes. There was also observed a kind of breakdown of the hydrodynamic regime. It consists in the duplication of the shear, purely imaginary modes, which  acquire non-vanishing real parts at some critical value of $\alpha$ \cite{Grozdanov:2016vgg,Grozdanov:2016fkt}. These modes may interfere  with the hydrodynamic modes in some range of parameters. However, in \cite{Konoplya:2017ymp} it was shown that the effect of acquiring non-zero real part occurs already in the region of gravitational instability, so that it is not clear whether this effect may actually have some holographic meaning.

It is natural to learn whether the above new features are artifacts of the particular gravitational backgrounds or appropriate to a broader class of models or, possibly, even universal for a broad class of theories with higher curvature corrections. The straightforward task is to extend the above analysis from planar to spherically symmetric black holes. Although by now there are a lot of papers on quasinormal modes of black holes and branes in Gauss-Bonnet and Lovelock theories, quasinormal modes of gravitational perturbations of Gauss-Bonnet-AdS black holes have not been studied so far. Therefore, here we shall analyze quasinormal modes of spherically symmetric Einstein-Gauss-Bonnet-anti-de Sitter black holes  in detail. In our previous paper \cite{Konoplya:2017ymp} it was shown that Gauss-Bonnet-AdS black holes can be unstable and the exact parametric region of instability was found. This instability, counter-intuitively, develops at high multipole numbers $\ell$ and was called therefore \emph{eikonal instability}. It is intrinsically related to the breakdown of the well-posedness of the initial value problem \cite{Reall:2014pwa}. A similar instability was found for asymptotically flat \cite{Gleiser:2005ra,Konoplya:2008ix} and de Sitter black holes \cite{Cuyubamba:2016cug} in Gauss-Bonnet theories, as well as for planar Einstein-Lovelock \cite{Takahashi:2010gz}, spherical pure Lovelock (i.e. without the Einstein term) black holes \cite{Gannouji:2013eka}, and for small charged Lovelock black holes \cite{Takahashi:2012np}. The full analysis of the eikonal instability of black holes and branes in the generic Lovelock theory has been recently performed in \cite{Konoplya:2017lhs}. However, test fields in the background of the Einstein-Gauss-Bonnet black hole do not show such an instability \cite{Konoplya:2004xx,Gonzalez:2017gwa}. At the same time, when the non-zero positive $\Lambda$-term is turned on, there is another, non-eikonal gravitational instability \cite{Cuyubamba:2016cug} of Gauss-Bonnet black holes, which may be similar in nature with the gravitational instability of higher dimensional Ressiner-Nordstr\"om-de Sitter black holes in Einstein gravity \cite{Konoplya:2008au}. The search for eikonal instability can be performed by analysis of the dominant (at high multipole numbers $\ell$) term of the effective potential only, so that other types of instability at lower $\ell$ cannot be excluded.
Thus, further detailed investigation of the quasinormal spectrum of Gauss-Bonnet-AdS black holes is also motivated, as it allows one to exclude or detect potential non-eikonal instabilities owing to non-zero (negative) $\Lambda$-term.

In the present paper we show that the spectrum consists of two essentially different, relatively the Gauss-Bonnet coupling $\alpha$, types of modes. First type of modes approaches Schwarzschild-AdS values when $\alpha \to 0$. In particular, the most interesting for the holography \emph{hydrodynamic} and \emph{sound} modes, can be expressed through their Einsteinian limits by adding a linear (in $\alpha$) correction. The analytical form of the correction is found by fitting the numerical data. The second type of modes are purely imaginary and their damping rates increase, approaching infinity when $\alpha \rightarrow 0$, i.e. these modes are non-perturbative  in $\alpha$ and do not exist in the Einsteinian (Schwarzschild-AdS) limit. In the regime of infinitely large, relatively the AdS radius, black holes we reproduce the black brane regime and discuss parametric region of instability of black branes and properties of quasinormal modes/thermalization spectra for stable region and near the threshold of instability. This way we also complement recent study of the black brane's spectrum in  \cite{Grozdanov:2016vgg,Grozdanov:2016fkt}.

The paper is organized as follows. Sec.~2 gives the basic information on the background black-hole metric, perturbation equations and diversity in terminology and designations between AdS/CFT and gravitational communities. Sec.~3 is devoted to analysis of the form of the effective potentials and to the search of a potential non-eikonal instability. Sec.~4 relates the details of the shooting method, which we used for finding the quasinormal modes. Sec.~5~and~6 discuss the found quasinormal modes for large and intermediate/small black holes respectively. Sec. ~7 discusses the timescale for relaxation of perturbations. In sec.~8 we summarize the obtained results and mention prospects for future investigation.

\section{The background metric and perturbation equations}

The Lagrangian of the D-dimensional Einstein-Gauss-Bonnet theory has the form:
\begin{equation}\label{gbg3}
  \mathcal{L}=-2\Lambda+R+\frac{\alpha}{2}(R_{\mu\nu\lambda\sigma}R^{\mu\nu\lambda\sigma}-4\,R_{\mu\nu}R^{\mu\nu}+R^2).
\end{equation}
The Gauss-Bonnet term corresponds to the full divergence in $D=4$, so that it contributes only to the higher dimensional space-times.
For $D>6$ there are higher than the second-order curvature corrections given by the Lovelock theory \cite{Lovelock:1971yv}.

An exact solution for a static spherically symmetric black hole in the $D$-dimensional Einstein-Gauss-Bonnet theory (\ref{gbg3}) was found by  Boulware and Deser in \cite{Boulware:1985wk}. The metric has the form:
\begin{equation}\label{gbg4}
 ds^2=-f(r)dt^2+\frac{1}{f(r)}dr^2 + r^2\,d\Omega_n^2,
\end{equation}
where $d\Omega_n^2$ is a line element of the $(n=D-2)$-dimensional sphere, and
\begin{equation}\label{fdef}
f(r)=1-r^2\,\psi(r),
\end{equation}
such that it satisfies the following relation:
\begin{equation}\label{Wdef}
W[\psi]\equiv\frac{n}{2}\psi(1 + \a\psi) - \frac{\Lambda}{n + 1} = \frac{\mu}{r^{n + 1}}\,.
\end{equation}
Here $\Lambda$ is a cosmological constant, $\mu$ is a constant, proportional to mass, and the Gauss-Bonnet coupling constant is
$$\a\equiv \alpha\frac{(n - 1) (n - 2)}{2}.$$
The black hole solution of (\ref{Wdef}), which goes over into the known Tangherlini solutions \cite{Tangherlini:1963bw} allowing, in general, for a non-zero $\Lambda$-term, is
\begin{equation}\label{psidef}
  \psi(r)=\frac{4\left(\frac{\mu}{r^{n+1}}+\frac{\Lambda}{n+1}\right)}{n+\sqrt{n^2+8\a n\left(\frac{\mu}{r^{n+1}}+\frac{\Lambda}{n+1}\right)}}.
\end{equation}
First of all, we are interested in this branch of solutions, because it has the known asymptotically flat, de Sitter, and anti-de Sitter analogues\footnote{In the general case, even when $\Lambda =0$, there is a branch of asymptotically anti-de Sitter solutions, which could also be studied.}. Let us describe the range of parameters corresponding to such a \emph{black hole} with the required AdS asymptotic at a negative $\Lambda$-term. As we shall have to investigate the spacetime behavior \emph{outside the black hole only}, it is useful to express the black hole mass in terms of its size by introducing the radius of the event horizon $r_H>0$. This choice comes at a price that the parametric space must be taken carefully: one must ensure that the black-hole mass is positive and the event horizon exists.

For a given negative value of $\a$, there is a lower bound on the mass parameter $\mu$:
\begin{equation}\label{massbound}
\mu>\frac{n\,(-2\a)^{(n-1)/2}}{4}\left(1+\frac{8\a\Lambda}{n(n+1)}\right),
\end{equation}
for which there is an event horizon $r_H>0$. From (\ref{massbound}) it follows that for $\a<0$ we have
\begin{equation}\label{horizonbound}
r_H^2>-2\a=-(n-1)(n-2)\alpha.
\end{equation}
The righthand side of (\ref{massbound}) is positive for any $\Lambda$, implying that the black hole exists only for a positive asymptotic mass.

In order to measure all the quantities in the units of the same dimension we express $\mu$ as a function of the event horizon $r_H$, as
\begin{equation}\label{massdef}
  \mu=\frac{n\,r_H^{n-1}}{2}\left(1+\frac{\a}{r_H^2}-\frac{2\Lambda  r_H^2}{n(n+1)}\right).
\end{equation}

Here we shall measure $\Lambda$ in units of the AdS radius $R$, defined by relation
\begin{equation}\label{Rdef}
\lim_{r\to\infty}\psi(r)=-\frac{1}{R^{2}}.
\end{equation}
Then, we have
\begin{equation}\label{AdSlambda}
  \Lambda=-\frac{n(n+1)}{2R^2}\left(1-\frac{\a}{R^2}\right),
\end{equation}
implying that
\begin{equation}
\a<R^2.
\end{equation}
It turns out that, when $R^2/2<\a<R^2$, the solution (\ref{psidef}) does not satisfy eq. (\ref{Rdef}) and describes a black hole, which is identical to the one with $\a<R^2/2$ after some re-scaling of parameters. Therefore, here we shall consider black holes with $\a \leq R^2/2$ only.
Other basic properties of Einstein-Gauss-Bonnet-AdS black holes were considered in \cite{Cai:2001dz}.

After decoupling of the angular variables the perturbation equations can be reduced to the second-order master differential equations \cite{Takahashi:2010ye}
\begin{equation}\label{wavelike}
\left(\frac{\partial^2}{\partial t^2}-\frac{\partial^2}{\partial r_*^2}+V_i(r_*)\right)\Psi(t,r_*)=0,
\end{equation}
where $r_*$ is the tortoise coordinate,
\begin{equation}
dr_*\equiv \frac{dr}{f(r)}=\frac{dr}{1-r^2\psi(r)},
\end{equation}
and $i$ stands for $t$ (\emph{tensor}), $v$ (\emph{vector}), and $s$ (\emph{scalar}) perturbations.
The explicit forms of the effective potentials $V_s(r)$, $V_v(r)$, and $V_t(r)$ \cite{Cuyubamba:2016cug} are given by
\begin{eqnarray}\nonumber
V_t(r)&=&\frac{\ell(\ell+n-1)f(r)T''(r)}{(n-2)rT'(r)}+\frac{1}{R(r)}\frac{d^2R(r)}{dr_*^2},\\\label{potentials}
V_v(r)&=&\frac{(\ell-1)(\ell+n)f(r)T'(r)}{(n-1)rT(r)}+R(r)\frac{d^2}{dr_*^2}\Biggr(\frac{1}{R(r)}\Biggr),\\\nonumber
V_s(r)&=&\frac{2\ell(\ell+n-1)f(r)P'(r)}{nrP(r)}+\frac{P(r)}{r}\frac{d^2}{dr_*^2}\left(\frac{r}{P(r)}\right),
\end{eqnarray}
where $\ell=2,3,4,\ldots$ is the multipole number and functions $T(r)$ and $R(r)$ can be written as follows
\begin{eqnarray}
T(r)&=& r^{n-1}\frac{dW}{d\psi}=\frac{nr^{n-1}}{2}\Biggr(1+2\a\psi(r)\Biggr),\\\nonumber
R(r)&=&r\sqrt{T'(r)},
\qquad P(r)=\frac{2(\ell-1)(\ell+n)-nr^3\psi'(r)}{\sqrt{T'(r)}}T(r).
\end{eqnarray}

\TABLE{
\begin{tabular}{|c|c|}
  \hline
 designations used in gravity & \emph{designations used in holography}  \\
  \hline
  scalar & \emph{sound} \\
    \hline
  vector & \emph{shear} \\
    \hline
  tensor & \emph{scalar} \\
  \hline
\end{tabular}
\caption{Terms used for types of gravitational perturbations in AdS/CFT- and gravity- oriented papers.}\label{Table1}
}

In the fields of gravity and holography different terms for the types of gravitational perturbations are used. Gravitationists work mostly with spherically symmetric black holes and are used to distinguish the three channels of perturbations according to the irreducible representations of the rotation group on $(D-2)$-sphere. Thus, harmonics are called \texttt{scalar}, \texttt{vecto}r, and \texttt{tenso}r. Holographic community analyzes mostly black holes with planar horizons and, following the hydrodynamic analogies, calls the corresponding channels: \emph{sound}, \emph{shear} and \emph{scalar} (see Table \ref{Table1}).

The Gauss-Bonnet coupling $\lambda_{GB}$ used usually in holography, and the coupling $\alpha$ used in gravity are related as follows:
\begin{eqnarray}\label{designations}
\lambda_{GB} \equiv\frac{\alpha}{L^2}\equiv-\frac{2\Lambda\alpha}{n(n+1)}&=&\frac{\alpha}{R^2}\left (1- \frac{2\alpha}{(n-1)(n-2)R^2}\right)\\\nonumber&&=\frac{2\a}{(n-1)(n-2)R^2}\left (1- \frac{\a}{R^2}\right).
\end{eqnarray}

In the regime of large, relatively the AdS-radius $R$, spherical black holes, the results for planar black holes (black branes) must be reproduced. For instance, the quasinormal modes of large AdS black holes found in \cite{Horowitz:1999jd} coincide with those of black branes \cite{Starinets:2002br} after the proper rescaling. Therefore, in order to compare the numerical data, it is necessary to know the relation between the multipole number $\ell$ in spherical symmetry and momentum $k$ in the planar one:
\begin{equation}
k \sim \frac{\ell}{r_{H}}, \quad r_{H} \to \infty.
\end{equation}

One can easily check that, in this limit formulas (44), (51),  and (53) of \cite{Gonzalez:2017gwa}, obtained for black holes, coincide, respectively, with (2.79), (2.80), and (2.81) of \cite{Grozdanov:2016fkt} for black branes.

\section{The effective potentials and stability}

The exact regions of the eikonal instability of the Einstein-Gauss-Bonnet-AdS black holes were found in \cite{Konoplya:2017ymp} (see fig.~1~and~2 therein). These regions were found by analysis of regime of dominance of a negative gap in the effective potential at high $\ell$. Thus, strictly speaking, it is not guaranteed that no instability exists for lower $\ell$, especially, taking into consideration that such non-eikonal instability at the lowest multipole $\ell=2$ was found for the Gauss-Bonnet-de Sitter black holes \cite{Cuyubamba:2016cug}.
Therefore, here we shall start from the analysis of positiveness of the effective potentials at $\ell=2$, concentrating on those regions which are free from the eikonal instability. If the effective potentials are positive definite everywhere outside the event horizon, then the black hole is stable, because the differential operator
\begin{equation}
-\frac{\partial^2}{\partial r^{*2}} + V_{eff}
\end{equation}
is a positive-definite self-adjoint operator in the Hilbert space of
square integrable functions $L(r^{*}, d r^{*})$, i.~e. there are no negative-mode solutions for well-behaved initial data (smooth data of compact support). In other words, all solutions are bounded all of the time. In the parametric regions of the eikonal instability, no such well-behaved initial data can be provided, because of absence of convergence in $\ell$. Nevertheless, the well-posed initial value problem is expected when there is no eikonal instability. We construct effective potentials and investigate regions of their positiveness for these eikonal-instability-free cases.

Thus, for example, the region for the eikonal instability  (see eqs. (4.3) and (4.5) in \cite{Konoplya:2017ymp}) in the scalar sector for $D=5$ is
\begin{equation}\label{uplimit}
\a>\frac{R^2r_H^2}{2}\frac{\sqrt{2}-1}{\sqrt{2}r_H^2+R^2},
\end{equation}
and in the tensor channel is
\begin{equation}\label{lowlimit}
\a<-\frac{R^2r_H^2}{2}\frac{\sqrt{3}-\sqrt{2}}{\sqrt{3}R^2+\sqrt{2}r_H^2}.
\end{equation}

For $D=6$ the region of instability obeys the following inequality (eq. 4.6 in \cite{Konoplya:2017ymp}):
\begin{eqnarray}\label{n4t}
&&4 \a^4 + 8 \a^3 r_H^2 - 44 \a^2 r_H^4 - 48 \a r_H^6  
- 6 r_H^8 - 20 \a^3 r_H^4 \Lambda - 20 \a^2 r_H^6 \Lambda + \a^2 r_H^8 \Lambda^2>0.
\end{eqnarray}

Construction of effective potentials for the lowest multipole numbers $\ell$ shows that the scalar and tensor $D=5$ potentials are positive-definite, while the vector potential has a negative gap near the event horizon for some values of the parameters. This gap does not lead to the eikonal instability, because here higher $\ell$ stabilize the system. At the same time, the vector potential can be easily deformed in such a way that the negative gap disappears, while the minimal allowed value of the damping rate of quasinormal frequencies does not become lower. This procedure is called S-deformation and was used in a number of works, devoted to analysis of stability of various black holes and fields \cite{Ishibashi:2003ap}. The S-deformation for Gauss-Bonnet black holes was done in \cite{Takahashi:2010gz} implying, first of all, asymptotically flat black holes, but the whole procedure is the same for $f(r)$ with any asymptotic behavior. Thus, according to  \cite{Takahashi:2010gz}, vector perturbations of Guass-Bonnet-(A)dS black holes are stable. Summarizing the $D=5$ case: \emph{$D=5$ Gauss-Bonnet-AdS black holes do not have other instability than the eikonal one, whose parametric region was found in \cite{Konoplya:2017ymp}}.

The $D=6$ case is different, because the scalar potential has a big negative gap, which is the deeper, the larger $r_{H}$ is.
In this case we need to make an extensive search for quasinormal modes in the region of parameters, which provides negativeness of the potential, and make sure that no growing modes appear in the spectrum. Unstable modes of a spherically symmetric background do not oscillate, as it was shown in \cite{Konoplya:2008yy}, so that, we need to search for unstable modes only on the imaginary axis. A restriction upon the possible values of unstable modes comes from the depth of the negative potential gap: since $V - \omega^2  > 0$ guarantees stability, $$Im(\omega)<\sqrt{-V_{min}},$$ where $V_{min}$ is the minimal value of the effective potential at $r_H\leq r<\infty$.
With the help of the shooting method, which will be described in the next section, we found no growing purely imaginary modes in the above region. This situation is not new for asymptotically anti-de Sitter space-times, as, for example, the effective potential of the Reissner-Nordstr\"om-AdS black hole has large and even infinite region of negativeness of the effective potential, but, owing to the asymptotic non-flatness of spacetime and specific behavior of $r^{*}$ coordinate, no instability develops in this case \cite{Konoplya:2008rq}.

\section{The shooting method}

In order to analyse gravitational stability of a black hole whose perturbations are governed by cumbersome effective potentials, numerical analysis of black holes' quasinormal spectrum is the only feasible way. For asymptotically AdS black holes, the quasinormal mode boundary conditions for gravitational perturbations require purely incoming waves at the event horizon and the Dirichlet boundary conditions at infinity. If a growing mode is found, then the considered system is unstable. Although, usually, damped quasinormal modes have both real and imaginary parts, i.~e. are oscillating, the growing modes for spherically symmetric background are known to be \emph{non-oscillating}, that is, \emph{pure imaginary} \cite{Konoplya:2008yy}. Thus, it is sufficient to look for unstable modes only along the imaginary axis.

The shooting method is useful for finding of quasinormal modes of asymptotically Gauss-Bonnet-AdS black holes. Its general features are related in \cite{Konoplya:2011qq}, while here, in a similar fashion with \cite{Konoplya:2008rq}, we shall adopt it for asymptotically anti-de Sitter space-times. Equation (\ref{wavelike}) has at least two regular singular points: at spacial infinity and at the event horizon. At the event horizon $V_i(r)\propto f(r)\propto (r-r_H)$, so that we have
$$\Psi(r)\propto(r-r_H)^{\pm\imo\omega/f'(r_H)}.$$
The quasinormal boundary conditions at the event horizon imply that
$$\Psi(r)=(r-r_H)^{-\imo\omega/f'(r_H)}\left(Z_0+{\cal O}(r-r_H)\right).$$
At spatial infinity the two linear independent solutions of $\Psi(r)$ are
\begin{eqnarray}
\Psi_1(r)\sim r^{-(D-4)/2}, &\Psi_2(r)\sim r^{(D-6)/2}, & D\neq5,\label{gen-infinity}\\\nonumber
\Psi_1(r)\sim r^{-1/2}, & \Psi_2(r)\sim r^{-1/2}\ln(r), & D=5,
\end{eqnarray}
The Dirichlet boundary conditions imply that (for $D>4$)
\begin{equation}\label{infinity-BC}
\Psi(r\to\infty)\sim r^{-(D-4)/2}.
\end{equation}

Let us consider the new function
\begin{equation}
y(r)=\left(1-\frac{r_H}{r}\right)^{\imo\omega/f'(r_H)}\Psi(r).
\end{equation}
If $\Psi(r)$ satisfies the quasinormal boundary conditions, then $y(r_H)=Z_0$. Since $y(r)$ satisfies the linear equation, one can choose $Z_0=1$ to fix the scale. Then $y'(r_H)$ can be found from equation (\ref{wavelike}) in the following way
\begin{equation}\label{horizoncond}
y'(r_H)=\frac{\imo\omega f''(r_H)}{2f'(r_H)^2}-\frac{\imo\omega}{r_Hf'(r_H)}+\frac{1}{f'(r_H)-2\imo\omega}\cdot\lim_{r\to r_H}\left(\frac{V_i(r)}{f(r)}\right).
\end{equation}
Using these initial conditions at the event horizon, we solve equation (\ref{wavelike}) numerically for each $\omega$ with the help of the \emph{Wolfram~Mathematica\textregistered} built-in function $NDSolve$ for $r_H\leq r \leq r_f$, where $r_f\gg r_H$ (the typical value of $r_f$ is $\sim 10^4r_H$).

In the general case the behavior of $\Psi(r)$ at infinity is a superposition of the two solutions (\ref{gen-infinity}), $\Psi_C(r)$ and $\Psi_D(r)$:
\begin{equation}\label{fit-function}
\Psi(r\to\infty)=Z_C \Psi_C(r)+Z_D\Psi_D(r),
\end{equation}
where $\Psi_C(r)$ satisfies the quasinormal boundary condition (\ref{infinity-BC}). If $\omega$ is the quasinormal frequency, then the corresponding solution must satisfy the boundary conditions (\ref{infinity-BC}) at spatial infinity and, thereby, $Z_D=0$.

In order to calculate $Z_C$ and $Z_D$ we obtain expansions of $\Psi_C(r)$ and $\Psi_D(r)$ far from the black hole in analytic form. The expansion for $\Psi_C(r)$ contains only negative powers of $r$,
$$\Psi_C(r)=r^{-\frac{D-4}{2}}\left(1+\frac{C_1}{r}+\frac{C_2}{r^2}+\frac{C_3}{r^3}\ldots\right).$$
Since the difference between the roots of the indicial equation is integer, the expansion for $\Psi_D(r)$ contains not only powers of $r$, but also contributions, proportional to $\ln(r)$. In particular, for $D=5$, one has
\begin{equation}\label{seriesD5} \Psi_D(r)=r^{-1/2}\ln(r)\left(1+\frac{A_1}{r}+\frac{A_2}{r^2}+\frac{A_3}{r^3}\ldots\right)+r^{-1/2}\left(\frac{B_1}{r}+\frac{B_2}{r^2}+\frac{B_3}{r^3}\ldots\right),
\end{equation}
while for $D=6$,
\begin{equation}\label{seriesD6}
\Psi_D(r)=1+\frac{\ln(r)}{r}\left(\frac{A_1}{r}+\frac{A_2}{r^2}+\frac{A_3}{r^3}\ldots\right)+\frac{1}{r}\left(\frac{B_1}{r}+\frac{B_2}{r^2}+\frac{B_3}{r^3}\ldots\right).
\end{equation}
As the series (\ref{seriesD5}, \ref{seriesD6}) are convergent, we have used only the first three terms of the expansions. The coefficients $A_1$, $A_2$, $A_3$, $B_1$, $B_2$, $B_3$, $C_1$, $C_2$, and $C_3$ are determined by substituting the expansions (\ref{seriesD5}, \ref{seriesD6}) into (\ref{wavelike}).

Thus, our numerical procedure is the following. For any given value of $\omega$ we integrate the equation (\ref{wavelike}) numerically, by imposing quasinormal boundary condition at the event horizon (\ref{horizoncond}). At a large distance from the black hole, near $r_f$, we fit the numerical values of $\Psi(r)$ by the analytical expression (\ref{fit-function}) and find $Z_C$ and $Z_D$ by solving the least squares problem. The quasinormal modes correspond to the roots of the equation
\begin{equation}\label{QNM-equation}
Z_D(\omega)=0.
\end{equation}

One should be careful when applying the shooting method to finding of higher overtones or situations when the purely imaginary mode acquires a non-zero real part, because in these cases the shooting method requires extra precision in order to reproduce the known analytical results accurately \cite{Gonzalez:2017gwa}.

\section{Large ($r_H/R \gg 1$) black holes}

Once the regions of instability are clearly determined, we are in position to analyze quasinormal modes in the stability sector. There are three channels of perturbations (scalar, vector and tensor) and two qualitatively different regimes of black holes: large and small (relatively the anti-de Sitter radius $R$). We shall discuss the features of the quasinormal spectrum for all these cases, concentrating on the most interesting from the point of view of the holography issues: the hydrodynamic and sound modes of large black holes.

\subsection{Scalar (\emph{sound}) channel}

\TABLE{
\begin{tabular}{|l|c|c|c|}
  \hline
  $\alpha/R^2$ & $\omega_{0}R$ & $\omega_{1}R$ & $\omega_{h}R$  \\
 \hline
  -0.1 & 1.6647-0.1805\imo & 21.2801-15.7602\imo & -28.3988\imo  \\ 
  \hline
 -0.09 & 1.6633-0.1761\imo  & 21.0643-15.9414\imo &   -30.3375\imo \\ 
   \hline
 -0.07 & 1.6606-0.1673\imo  & 20.5849-16.2473\imo &   -35.6491\imo \\ 
   \hline
 -0.065 & 1.6600-0.1652\imo & 20.4578-16.3080\imo &   -37.6820\imo \\ 
   \hline
 -0.05 & 1.6580-0.1587\imo & 20.0715-16.4434\imo & c  \\
   \hline
 ~0 & 1.6517-0.1376\imo & 19.0355-16.3915\imo & --  \\
    \hline
 ~0.05 & 1.6460-0.1173\imo & 18.6983-16.2046\imo & c  \\
    \hline
 ~0.1 & 1.6408-0.0979\imo & 18.9832-16.1159\imo  & c  \\
    \hline
\end{tabular}
\caption{Quasinormal modes of the Gauss-Bonnet-AdS black hole: $D=5$, scalar channel, $\ell =2$, $r_{H}/R =6$. The frequencies are: $\omega_{0}$ is the fundamental sound mode, $\omega_1$ is the branch of modes with spacing linear in $r_{H}/R$, $\omega_{h}$ is the lowest mode on the imaginary axis.}\label{Table2}
}

There are three essentially different kinds of modes in the scalar channel (see table~\ref{Table2}):
\begin{enumerate}
\item perturbative in $\alpha$, \emph{sound mode}, which usually has the lowest damping rate and whose real part represents the speed of sound in a medium ($\omega_0$ in Table~\ref{Table2});
\item perturbative (in $\alpha$) branch of modes which scale as $r_{H}/R$  and go over into their Schwarzschild-AdS limits \cite{Konoplya:2003dd} when $\alpha =0$. The dominant mode of this series is shown as $\omega_1$ in Table~\ref{Table2}. At higher $n$ this branch of the spectrum is equidistant, as it takes place for Schwazrschild-AdS case. Though, detecting of high overtones $n$ is difficult by the shooting method.
\item branch of purely imaginary modes, which do not exist in the $\alpha =0$ limit, because they  ``come from infinity'' at small $\alpha$ along the negative imaginary axis, approaching and crossing the origin at the value of $\alpha$ corresponding to the threshold of eikonal instability. Thus, these modes are \emph{non-perturbative in $\alpha$}. Notice, that in Table (\ref{Table2}) data for some of the modes in the non-perturbative branch are absent, because at sufficiently small $\alpha$ the frequencies become too large and are difficult to detect by the shooting method.
\end{enumerate}

\FIGURE{
\includegraphics[width=0.5 \textwidth]{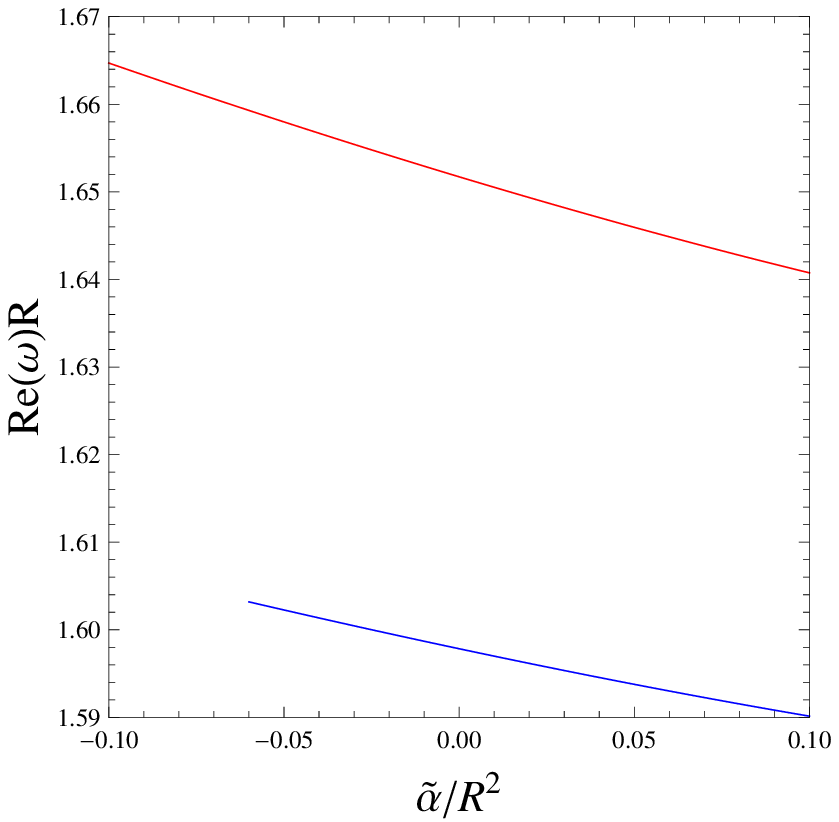}\includegraphics[width=0.5 \textwidth]{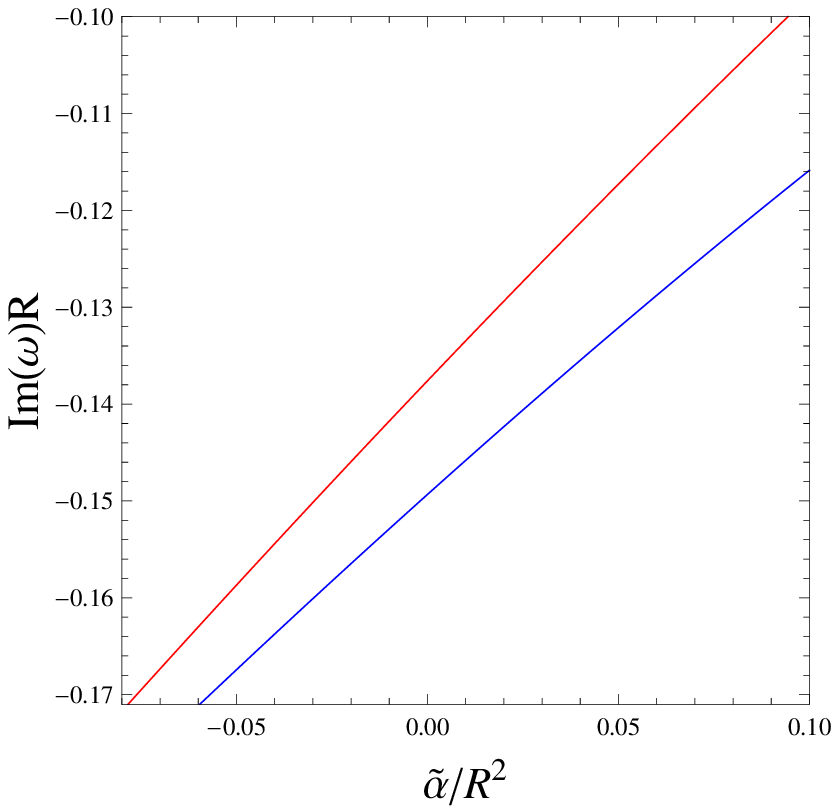}
\caption{Real (left panel) and imaginary (right panel) part of the fundamental scalar (sound) quasinormal modes for various values of $\a$ for $D=5$ (upper, red) and $D=6$ (lower, blue): $r_{H}/R = 6$, $\ell=2$. }\label{fig:sound}
}

The values of the sound mode given in fig.~\ref{fig:sound} for $\ell =2$, $r_{H}/R = 6$ can be very well fitted by the linear law:
\begin{eqnarray}
\omega R \approx 1.65209 (1 - 0.073 \a/R^2) - 0.138179 (1 -  2.99 \a/R^2) \imo, &\quad& D=5,\\
\omega R \approx 1.59797 (1 - 0.051 \a/R^2) - 0.149664 (1 -  2.31 \a/R^2) \imo, &\quad& D=6.
\end{eqnarray}

\FIGURE{
\includegraphics[width=\textwidth]{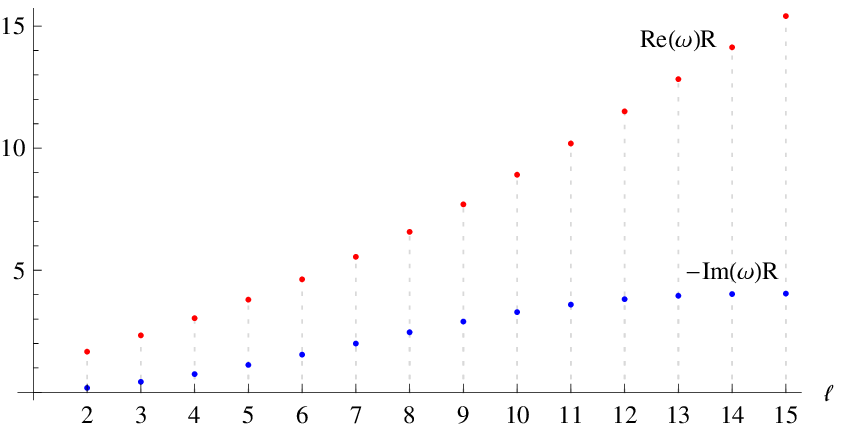}
\caption{Fundamental scalar (sound) quasinormal modes for various values of the multipole parameter $\ell$: $D=5$, $\alpha =-0.1R^2$, $r_{H}/R = 6$. }\label{fig:LargeL}
}

\FIGURE{
\includegraphics[width=0.5\textwidth]{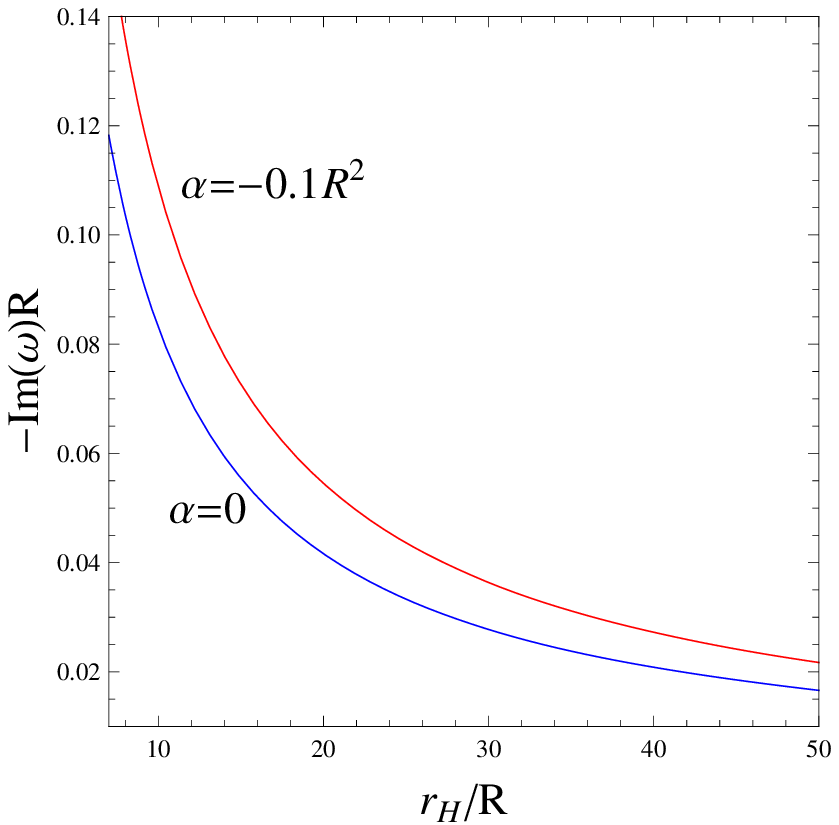}\includegraphics[width=0.5\textwidth]{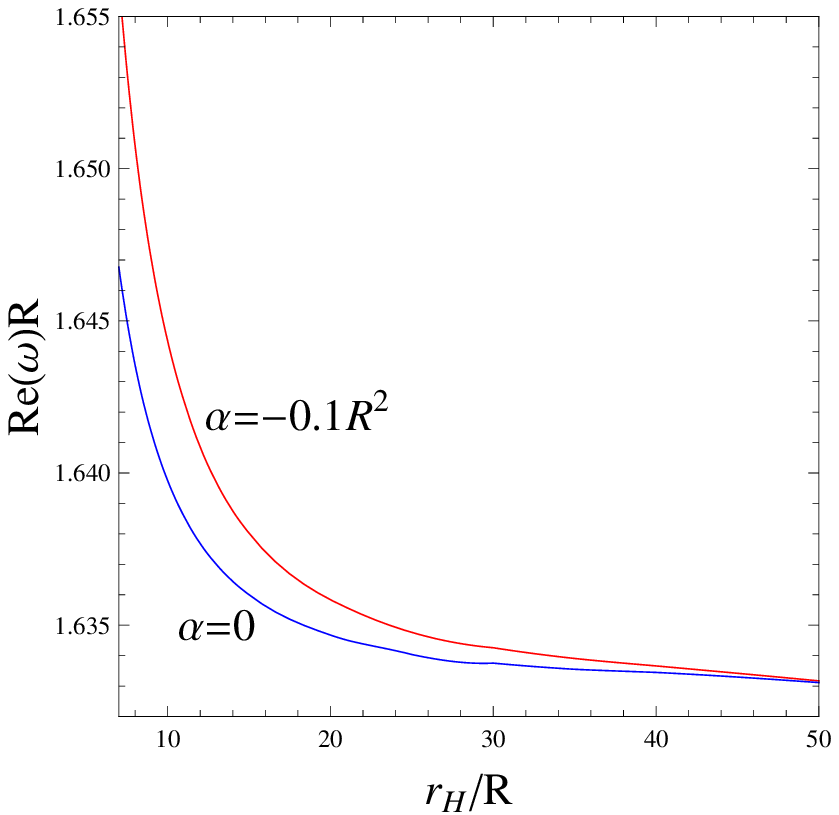}
\caption{$D=5$, $\alpha=0$ (blue) and $\alpha =-0.1R^2$ (red): sound mode ($\ell=2$) as a function of the black-hole radius. In the limit of large black holes $\Re{\omega}$ does not depend on $\alpha$ (right panel).}\label{fig:LargeRadius}
}

From fig.~\ref{fig:LargeL} one can see that, at a fixed $\alpha$, $\Re{\omega}$ of the sound mode is growing and at sufficiently large $\ell$ this growth looks linear. The damping rate, given by imaginary part of the frequency, approaches some constant. Notice, that when $\ell$ in fig.~\ref{fig:LargeL} is larger than $r_{H}/R$, such regime is irrelevant for holography, because it corresponds to a very large momentum of matter propagating on a relatively small sphere. In the other limit, of large black hole radius $r_{H}\gg R$, the real oscillation frequency does not depend on $\alpha$ as can be noticed from fig.~\ref{fig:LargeRadius}. Taking into consideration that at the leading order $\Re{\omega}$ of the sound mode equals $c_{s} k$ (where $c_s$ is speed of sound and $k$ is momentum), this may indicate that $c_{s}$ is unaffected by the Gauss-Bonnet coupling $\alpha$. Indeed, this is also confirmed by the calculations for black branes (see eq.~3.25 in~\cite{Grozdanov:2016fkt}).

By fitting a lot of numerical data, we can guess that when $\alpha$ is nonzero, the imaginary part of $\omega$ obeys the following relation:
\begin{eqnarray}\label{bbb}
\Im{\omega} = \Im{\omega_{SAdS}}\left(1 - A(D)\cdot\frac{\a}{R^2} \right) = - \frac{2}{3}\frac{(\ell -1)(\ell+D-2)}{(D-1) r_{H}} \left(1 - A(D)\cdot\frac{\a}{R^2} \right),
\end{eqnarray}
where $A(D) \approx 3$ for $D=5$, $A(D) \approx 2.3$ for $D=6$, $A(D) \approx 2$ for $D=7$, etc. Thus, we can suppose that $A(D) \approx (D+1)/(D-3)$. 
This relation is satisfied with good accuracy not only for very large, but also for moderately large (e.~g. $r_H/R = 6$) black holes. For sufficiently large black holes, approaching the black brane regime, the quasinormal frequencies are
\begin{eqnarray}\label{aaa}
\omega \approx \Re{\omega_{SAdS}} - \Im{\omega_{SAdS}} \left(1 - \frac{D+1}{D-3} \cdot \frac{\a}{R^2} \right)\imo, \quad r_{H} \gtrapprox R,
\end{eqnarray}
where $\omega_{SAdS}$ is the sound mode of the Schwarzschild-AdS black hole ($\alpha=0$).

\subsection{Vector (\emph{shear}) channel}

The vector channel has also three different types of modes:
\begin{itemize}
\item purely imaginary ``hydrodynamic'' mode;
\item purely imaginary non-perturbative in $\alpha$  modes (similar and numerically close to those found for the scalar channel);
\item modes with non-zero real part, which scale as $r_{H}/R$ and have the Einsteinian analogues in the limit $\alpha =0$ \cite{Konoplya:2003dd}. These frequencies are also numerically close to those of the scalar and tensor channels.
\end{itemize}

The vector modes allow one to compute the  $\eta/s$ ratio \cite{Son:2007vk,Brigante:2007nu}. The purely imaginary hydrodynamic modes for large Schwarzschild-anti-de Sitter black hole have the form (see formula~(16) in~\cite{Konoplya:2003dd})
\begin{equation}\label{SAdSlimit1}
\omega  = - \frac{(\ell -1)(\ell+D-2)}{(D-1) r_{H}} i, \quad r_{H} \gtrapprox R.
\end{equation}

In order to see how expression (\ref{SAdSlimit1}) is modified when adding the non-zero $\alpha$, we plotted the numerically found, purely imaginary mode $\omega$ as a function of $\alpha$ for various values of $\ell =2, 3, 4, 5$, $r_{H}/R =20$ (see fig.~\ref{fig:vector}).
The data fits very well the following formulas for $D=5$,
\begin{eqnarray}\nonumber
\omega R = -  0.063 \left(1 - 2.995 \frac{\a}{R^2}\right)\imo, &\quad& \ell =2,\\
\nonumber
\omega R = -  0.151 \left(1 - 3.002 \frac{\a}{R^2}\right)\imo, &\quad& \ell =3,\\
\nonumber
\omega R= -  0.264 \left(1 - 3.013 \frac{\a}{R^2}\right)\imo, &\quad& \ell =4,\\
\nonumber
\omega R = -  0.403 \left(1-  3.021 \frac{\a}{R^2}\right)\imo, &\quad& \ell =5,
\end{eqnarray}
and for $D=6$,
\begin{eqnarray}\nonumber
\omega R = -  0.060 \left(1- 2.290 \frac{\a}{R^2}\right)\imo, &\quad& \ell =2,\\
\nonumber
\omega R = -  0.140 \left(1- 2.293 \frac{\a}{R^2}\right)\imo, &\quad& \ell =3,\\
\nonumber
\omega R = -  0.241 \left(1- 2.297 \frac{\a}{R^2}\right)\imo, &\quad& \ell =4,\\
\nonumber
\omega R = -  0.361 \left(1- 2.301 \frac{\a}{R^2}\right)\imo, &\quad& \ell =5.
\end{eqnarray}

The factors in these fits approximate the SAdS expression (\ref{SAdSlimit1}). For example, for $r_{H}/R =20$, $\ell =5$ and $D=5$, the exact formula (\ref{SAdSlimit1}) gives $\omega R = -0.4\imo$,  while the numerical data is $\omega R = -0.403\imo$.
Thus, we conclude that when $\a$ is nonzero, formula (\ref{SAdSlimit1}) can be generalized as follows:
\begin{eqnarray}\label{QNMhydro}
\omega = - \frac{(\ell -1)(\ell+D-2)}{(D-1) r_{H}} \left(1 - \frac{D+1}{D-3}\cdot \frac{\a}{R^2}  \right)\imo, \quad \frac{r_{H}}{R} \gtrapprox \ell.
\end{eqnarray}

\FIGURE{
\includegraphics[width=0.5 \textwidth]{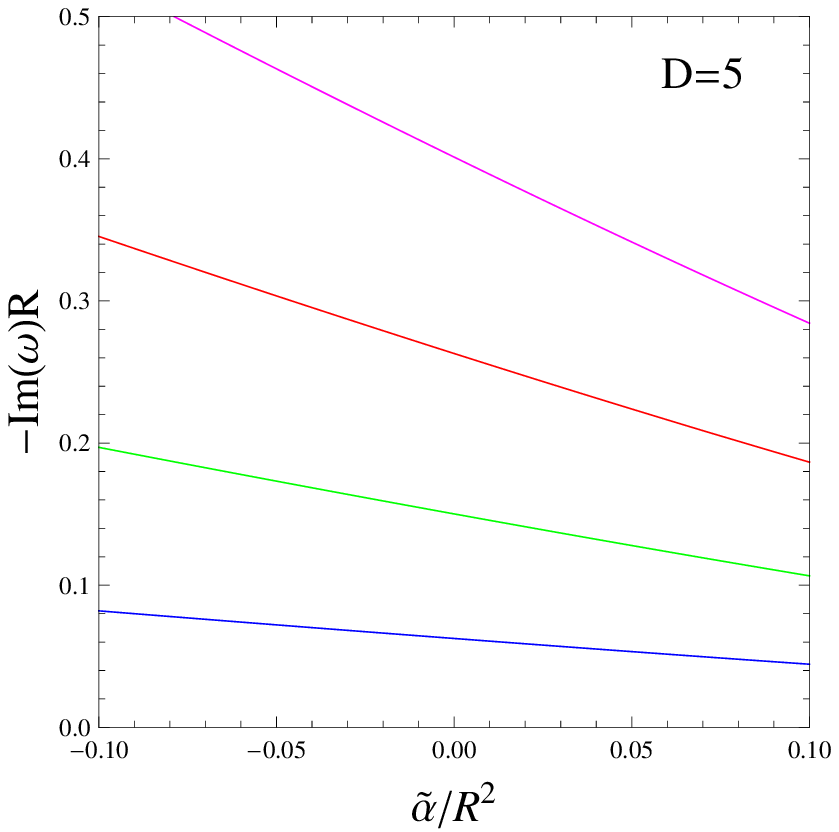}\includegraphics[width=0.5 \textwidth]{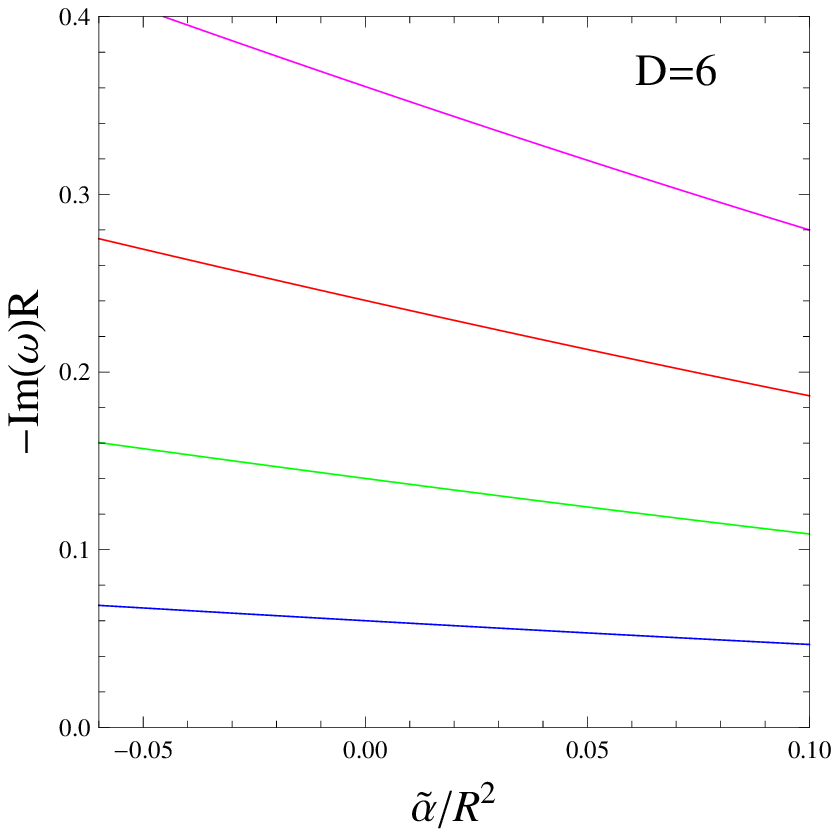}
\caption{Purely imaginary vector (shear) mode for $D=5$ (left panel) and $D=6$ (right panel) as a function of $\a$ for $r_{H}/R = 20$: $\ell =2$ (blue), $\ell =3$ (green), $\ell =4$ (red), $\ell =5$ (magenta). }\label{fig:vector}
}

\FIGURE{
\includegraphics[width=\textwidth]{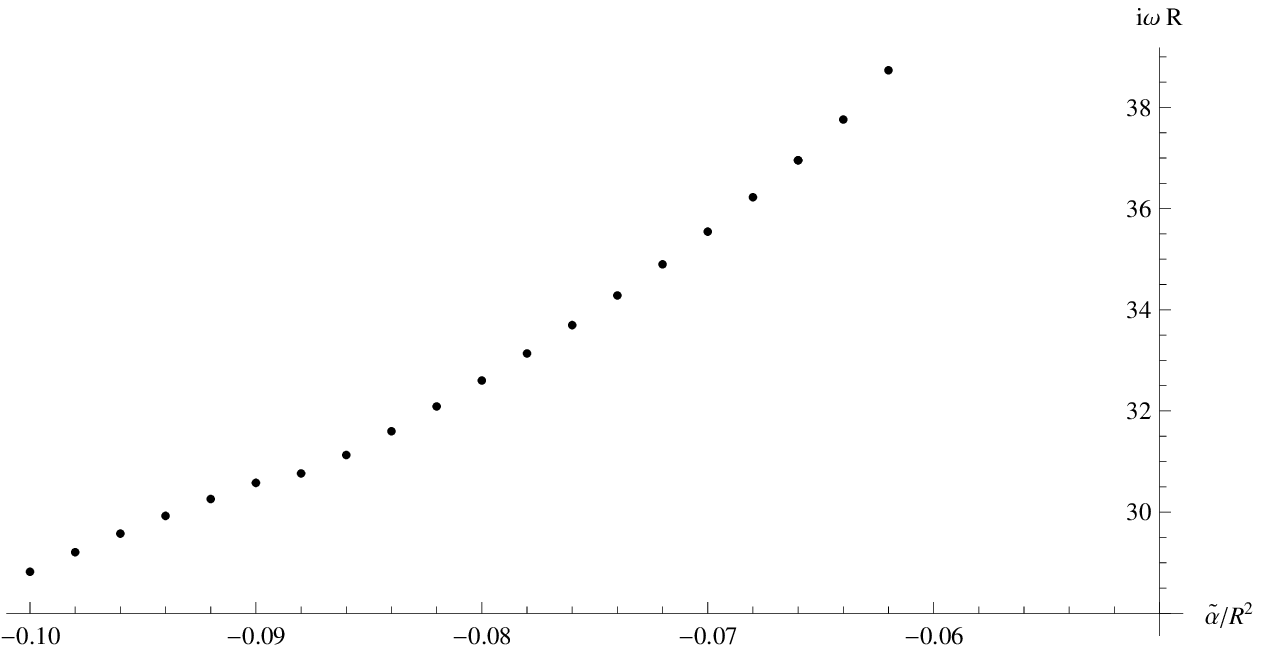}
\caption{Purely imaginary non-perturbative in $\a$ vector modes ($D=5$, $r_{H}/R = 6$) as a function of $\alpha$ for $\ell =2$. }\label{fig:VectorD5-non-pert}
}

Comparing the formulas for the sound and hydrodynamic modes (\ref{aaa}) and (\ref{QNMhydro}), we observe that the damping rates of these modes are related as follows:
\begin{equation}\label{ratio2}
\Im{\omega_{sound}} = \frac{2}{3}\cdot\Im{\omega_{hydro}}.
\end{equation}
The relation (\ref{ratio2}) also occurs for the Einstein-Gauss-Bonnet black branes \cite{Brigante:2007nu}.

The purely imaginary non-perturbative modes for all three types of gravitational perturbations have damping rate which increases as $\alpha$ is deceasing (fig.~\ref{fig:VectorD5-non-pert}). Thus, for sufficiently small $\alpha$ they reach arbitrarily large damping rates and their contribution to the spectrum is negligible, while at $\alpha =0$ they disappear from the spectrum.

At some sufficiently large $\alpha$ the purely imaginary, non-perturbative modes acquire the non-zero real part at some lower $\ell$. This occurs in the region of eikonal instability, so that higher $\ell$ modes are unboundedly growing \cite{Konoplya:2017ymp}. One can see how the purely imaginary ``mode'' becomes oscillating though the exact solutions of the wave equations found in \cite{Gonzalez:2017gwa} (see, for example eq. (28) therein) for the unstable case $\a=R^2/2$.

\subsection{Tensor (\emph{scalar}) channel}

\FIGURE{
\includegraphics[width=0.5\textwidth]{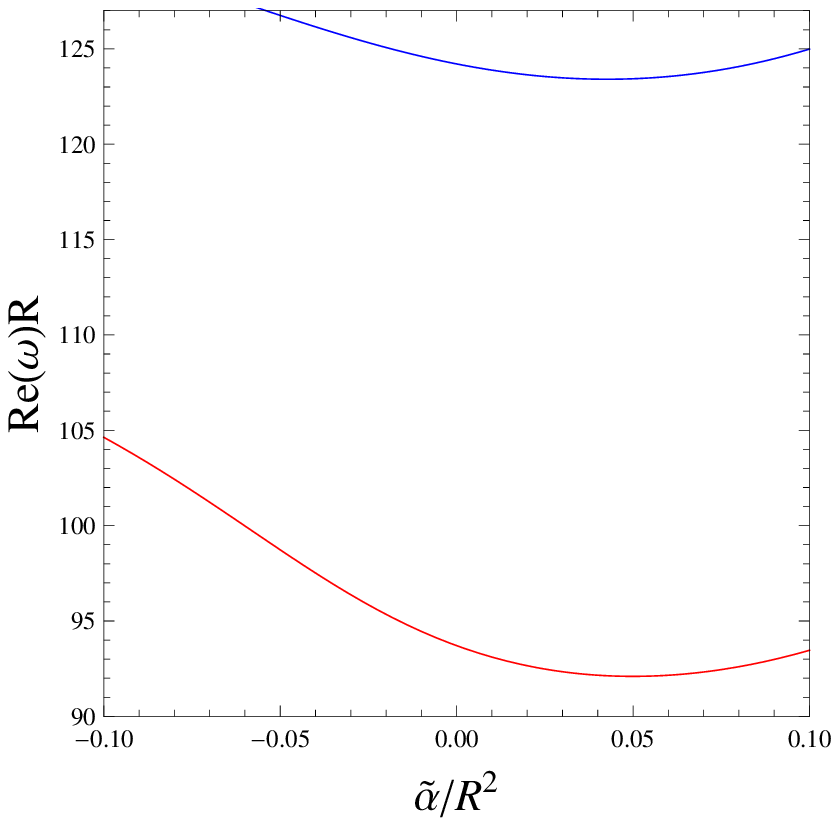}\includegraphics[width=0.5\textwidth]{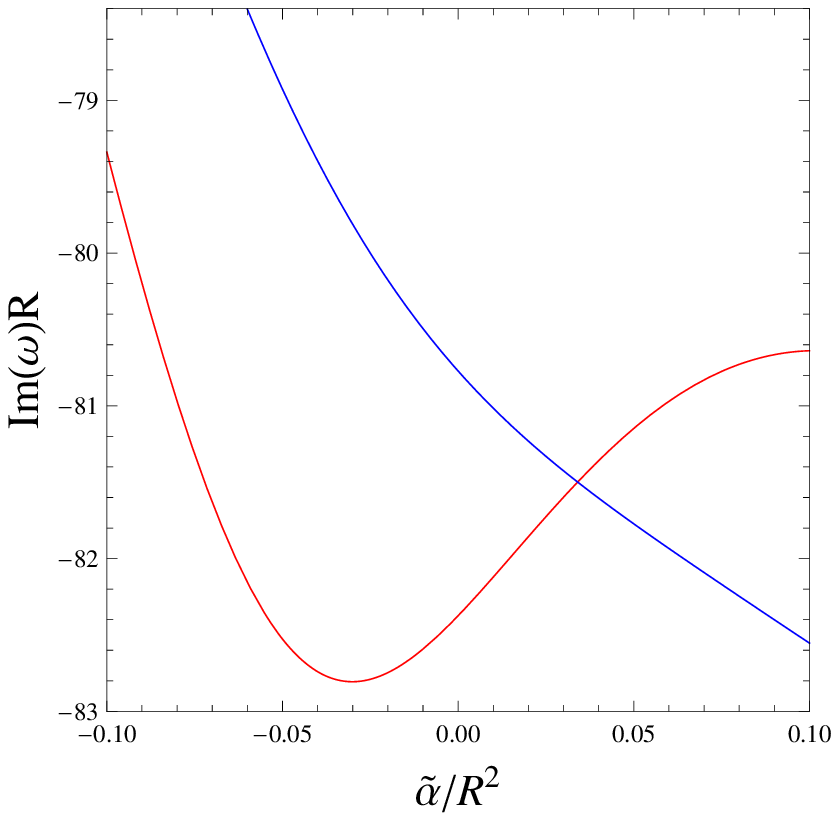}
\caption{Real (left panel) and imaginary (right panel) parts of $\omega$: $D=5$ (red) and $D=6$ (blue), $r_{H}/R = 30$, tensor channel, as a function of $\alpha$ for $\ell =2$. }\label{fig:tensor}
}

Tensor channel has only two kinds of modes: the same non-perturbative branch as for scalar and vector channels and the one scaling as  $r_{H}/R$ and having the Einsteinian limit when $\alpha \to 0$. Tensor perturbative modes do not lay on the linear fit, as can be seen from fig.~\ref{fig:tensor}, while the non-perturbative branch behaves similar with those for the other two channels and numerical values of the frequencies are close for all three types of perturbations. The non-linear behavior depicted on fig.~\ref{fig:tensor} is different from the linear in $\alpha$ sound and hydrodynamic modes. As we cannot completely exclude that the shooting method simply becomes inaccurate in this case\footnote{See also discussion on the discrepancy between the shooting method and analytical results in \cite{Gonzalez:2017gwa}.}, calculations with the help of an alternative method would be desirable.

\section{Moderate ($r_H/R \sim 1$) and small black holes}

Small black holes are not interesting from the point of view of possible applicability within AdS/CFT correspondence. Nevertheless, in order to have the full picture, we shall consider this regime as well.

Quasinormal modes of Schwarzschild-AdS black holes approach the normal modes of the empty AdS space-time when $r_{H} \to 0$ \cite{Konoplya:2002zu}. This is evidently not applicable to the Gauss-Bonnet case, because the limit $r_{H} \to 0$ is not achievable at any non-zero value of $\alpha$, since all sufficiently small black holes are unstable (see eqs.~\ref{uplimit},~\ref{lowlimit},~\ref{n4t}).

The eikonal instability develops due to the purely imaginary nonperturbative frequencies of the scalar or tensor channel. When the absolute value of $\alpha$ grows, the damping rates decrease for sufficiently large $\ell$. Then, \emph{at a fixed $\ell$} the non-perturbative frequency with the slowest decay rate becomes dominant and, at some critical value of $\alpha$, it turns into the unstable (growing) mode. When increasing $\ell$,  the critical value of $\alpha$ (at a given $\ell$) approaches the threshold of the eikonal instability \cite{Konoplya:2017ymp}, so that for any given $\alpha$ inside the region of instability there is always a finite value of the multipole number $\ell_{inst}$ for which a growing mode appears in the spectrum. The lowest unstable multipole number  can be arbitrarily large, if the value of $\alpha$ is sufficiently close to the threshold of the eikonal instability. While perturbations of higher than $\ell_{inst}$ multipoles are always unstable, perturbations of lower multipoles do not necessary contain a growing mode. Thus, in \cite{Konoplya:2017ymp} it was observed that within the region of eikonal instability the lower-multipole non-perturbative frequencies are damped and may ``duplicate'' by acquiring a real part (with positive and negative signs).


\section{Thermalization timescale}

According to the AdS/CFT correspondence, perturbation of a spherical, $D$-dimensional asymptotically AdS black hole is related to the perturbation of the thermal state propagating on the sphere in the $(D-1)$-dimensional world. When the black hole is sufficiently large, we approach the results of the black brane perturbations and the quantum system can be considered as the one not limited by a sphere. Damping rate of the dominant quasinormal mode characterizes the time, which is necessary for the dual quantum system to return to the thermal equilibrium. Thus, the thermalization timescales can be expressed as
\begin{equation}
\tau = \frac{1}{|Im (\omega)|}
\end{equation}
When the Gauss-Bonnet coupling is turned on, the Hawking temperature of large black holes can be approximated as follows:
\begin{equation}
T = \frac{f'(r_{H})}{4 \pi} =\frac{D-3}{4\pi r_H}+\frac{D-1}{4\pi r_H}\cdot\frac{R^2r_H^4-\a (r_H^4+R^4)}{R^4(r_H^2+2\a)}=\frac{(D-1)r_H}{4\pi R^2}\left(1-\frac{\a}{R^2}+{\cal O}(r_H^{-2})\right),
\end{equation}
being, thereby, linear in $\a$ and $r_{H}$. Therefore, using (\ref{aaa}) and (\ref{QNMhydro}) for the fundamental scalar (sound) and vector (hydrodynamic) modes, one has
$$\frac{\Im{\omega}}{T}=\frac{\Im{\omega_{SAdS}}}{T_{SAdS}}\left(1-\frac{4\a}{D-3}\frac{1}{R^2-\a}\right)+{\cal O}(r_H^{-2}).$$
Taking into account that $|\a|\lesssim0.1R^2$, we can neglect $\a$ in the denominator of the above expression, so that
$$\frac{\Im{\omega}}{T}\approx\frac{\Im{\omega_{SAdS}}}{T_{SAdS}}\left(1-\frac{4}{D-3}\frac{\a}{R^2}\right)=\frac{\Im{\omega_{SAdS}}}{T_{SAdS}}\left(1-\frac{2(D-4)\alpha}{R^2}\right).$$
The above relation connects the damping rates to the temperature ratios for large black holes in the Einstein and Einstein-Gauss-Bonnet theories. The sound mode has the longest lifetime, so that it determines the characteristic time for relaxation of perturbations.

An essential constrain on possible applicability of the Einstein-Gauss-Bonnet-AdS black hole and brane backgrounds must be imposed by the observed eikonal instability. If the GB coupling is large enough, black holes (and branes) are unstable \cite{Konoplya:2017ymp}, so that the black hole does not approach an equilibrium state, which could be characterized by a given set of black-hole parameters, Hawking temperature of the event horizon, etc. Thus, no relaxation of perturbations and thermalization should occur in such an unstable system.

In our opinion, when considering holographic applications one should not turn the blind eye to the eikonal instability, justifying it by the fact that the phenomena occurs at large momentum $k$, while one can be constrained by small momentum in the field theory side. First, when slightly increasing the Gauss-Bonnet coupling constant, the instability occurs at lower and lower momenta, reaching, for the large black hole case, the minimal $\ell=2$ multipole. For example, taking limit of $r_H\to\infty$ in (\ref{uplimit}) and ({\ref{lowlimit}}) and substituting into (\ref{designations}), we find that the five-dimensional black brane is gravitationally unstable outside the following region
\begin{equation}
-\frac{1}{8} \leq \lambda_{GB} \leq \frac{1}{8}.
\end{equation}
Once an instability takes place, then the following issue arises. The perturbation contains various values of momenta and in order to ignore the non-equilibration at high momenta, one should develop mathematically consistent cut-off for this case, if that is possible.

Thus, when discussing the possible holographic model for quantum liquids possessing superfluidity one can formally derive the $\alpha$-corrected formula for the viscosity to entropy density ratio \cite{Brigante:2007nu},
\begin{equation}
\frac{\eta}{s} \approx \frac{\hbar}{4 \pi k} (1- 4 \lambda_{GB}).
\end{equation}
In this context the regime $\lambda_{GB} \lessapprox 1/4$ in which the viscosity could be made small was considered in the literature \cite{Brigante:2007nu,Grozdanov:2015asa}. Here from (\ref{designations}) ($n=3$) we can see that the above regime corresponds to $\alpha \lessapprox R^2/2$, which is unstable. Thus, the holographic description of zero viscosity regime and the conclusions made within the Einstein-Gauss-Bonnet-AdS model look questionable and, should not be applied to quantum liquids with high momentum.

Another interesting question is related to the causality violation in theories with higher curvature corrections. For the black brane regime, the causality is violated for an evidently larger region ($\lambda_{GB}>9/100$) \cite{Brigante:2008gz} than the one suffering from instability ($\alpha > 0.146 R^2$, $\lambda_{GB} > 0.125$). At the same time the analysis of causality suggested in \cite{Brigante:2008gz} implies a well defined eikonal regime (see e.g. eq.~(26) in~\cite{Brigante:2008gz}), valid Fourier transformations (eq.~(7) in~\cite{Brigante:2008gz}), etc., in order to work with quasi-particles and show that they can travel faster than the speed of light.
All these constituents are not well defined when the eikonal instability occurs.

Recently, an analytical deduction of the viscosity to the entropy density ratio has been found for the system with translational symmetry by representing the black brane within a classical membrane paradigm \cite{Jacobson:2011dz}. It would be interesting to deduce the $\eta/\mu$ ratio from the quasinormal spectrum obtained here and compare it with that found in \cite{Jacobson:2011dz}.

\section{Discussion}

Here we have studied quasinormal modes of gravitational perturbations of the Einstein-Gauss-Bonnet-AdS black holes in detail.
The numerical analysis of the quasinormal spectrum has shown that there is no other than eikonal instability for such black holes.
The numerical data for the sound and hydrodynamic modes are shown to be very well described by an analytical formula with a linear in $\alpha$ correction to the corresponding modes of the Schwarzschild-AdS black hole. At the same time, when $\lambda_{GB}$ is not small enough, the black holes and branes suffer from the instability, so that the holographic interpretation of perturbation of such black holes become questionable, as, for example, the claimed viscosity bound violation \cite{Brigante:2007nu}. The eikonal instability phenomenon should be taken into account when considering perturbations of various black holes and branes in the Gauss-Bonnet theory \cite{Chakrabarti:2006ei}.

The non-perturbative in $\alpha$ quasinormal modes exist not only for considered here gravitational perturbations of the Einstein-Gauss-Bonnet-AdS black holes, but also in a number of other cases, which we would to like summarize here:
\begin{itemize}
\item Gravitational perturbations of asymptotically flat Einstein-Gauss-Bonnet black holes in various $D$\footnote{Therefore, asymptotically de Sitter black holes should also have such modes at least when $\Lambda$-term is sufficiently small. However, it was not checked in the literature so far. In the same manner, Lovelock theories also must have these modes at least when higher than second order in curvature coupling constants are small.} \cite{Konoplya:2008ix,Chen:2015fuf}.
\item Gravitational perturbations of $D=5$ asymptotically AdS black branes in the Einstein-Gauss-Bonnet and -$R^4$ theories \cite{Grozdanov:2016vgg}.
\item Gravitational perturbations of Einstein-Gauss-Bonnet-AdS spherical black holes (shown here and in \cite{Konoplya:2017ymp,Chen:2017hwm}).
\item Test scalar field perturbations in the background of Einstein-Gauss-Bonnet-AdS spherical black holes \cite{Gonzalez:2017gwa}.
\end{itemize}

All the above cases may indicate \emph{possible independence of existing of purely imaginary non-perturbative in $\alpha$ quasinormal modes on such peculiarities as: spin of a perturbed field, number of spacetime dimensions, asymptotic behavior (flat, dS, AdS), and, possibly, even character of higher curvature corrections (Gauss-Bonnet, Lovelock, $R^4$, etc.). } The broadness of this phenomena may have implications for $D=4$ large astrophysical black holes. Thus, it is reasonable to check the presence of the non-perturbative branch of modes for Einstein-dilaton-Gauss-Bonnet black holes, dynamical Chern-Simons black holes, various black hole solutions in f(R) gravity etc.
The gravitational spectra of the latter cases were investigated recently only from the point of view of small deviations from their Schwarzschild and Kerr values \cite{Cardoso:2009pk}, so that it seems reasonable to reconsider these works taking into account possible existence of non-perturbative modes.
As nowadays, gravitational wave experiments do not suggest a strict constrain on possible deviations from Kerr geometry \cite{Konoplya:2016pmh}, the coupling constants in higher curvature corrections can be sufficiently large, so that the non-perturbative modes have a chance to be dominating in the signal at some values of the parameters. These cases will be studied in our future publications \cite{workinprogress}.

Our work can be extended in a number of other ways: by adding an axion parameter \cite{Kuang:2017cgt}, considering higher curvature corrections and higher $D$.

\section*{Acknowledgments}

R.~K. would like to thank the Rudolf Peierls Centre for Theoretical Physics of University of Oxford for hospitality and partial support and the Bridging Grant of the University of T\"ubingen. A.~Z. thanks Conselho Nacional de Desenvolvimento Cient\'ifico e Tecnol\'ogico (CNPq) for support and Theoretical Astrophysics of Eberhard Karls University of T\"ubingen for hospitality. At its final stage this work was supported by ``Project for fostering collaboration in science, research and education'' funded by the Moravian-Silesian Region, Czech Republic and by the Research Centre for Theoretical Physics and Astrophysics, Faculty of Philosophy and Science of Sileasian University at Opava.

\end{document}